\documentclass[epj]{svjour}

\usepackage{graphicx}
\usepackage{amsmath}

\begin{document}
\title{Nonlinear electron emission under the action of ultrashort laser pulse}

\author{P.A. Golovinski\inst{1,2,}\thanks{\emph{Present address:} golovinski@bk.ru} \and E.A. Mikhin\inst{2}
}                     
\offprints{}          
\institute{Moscow Institute of Physics and Technology, Russia \and Physics Research Laboratory, Voronezh State Technical University, Russia}
\date{Received: }
%
\abstract{The process of nonlinear electron emission from a metal surface under the action of femtosecond laser pulse with moderate intensity $\sim10^{11}$~W/cm$^2$  is considered. One-dimensional model is formulated, taking into account the advantage of the $p$-polarized light in the nonlinear emission. The time dependent Schrodinger equation with fixed equilibrium boundary conditions is solved in the half-space using the Laplace transform technique. The energy spectrum of emitted electrons is obtained, and its dependence on the laser pulse parameters  in compare with experimental data is discussed.%
	} 
\maketitle
\section{Introduction}
\label{intro}

Field emission cathode has been developed to be an effective solution to produce electron beam, required for many applications including coherent radiation sources \cite{RefJ1,RefJ2} or as point sources of electrons for applications in high resolution electron microscopy \cite{RefJ3,RefJ4}. The field emission process is a type of electron emission, which involves the quantum tunneling of electrons from a surface to vacuum subjected to a strong applied electrical field. Cold field emission can produce electron beams in which a large current density is concentrated to a small solid angle, so such beams are said to be bright.  High current density is a consequence of the number of tunneling electrons being large if the field emission barrier is properly thin. For the barrier to be thin, dc fields on the order of GV/m are required at the surface.

One way of generating such fields is fabrication the surfaces with sharp nm-scale emitters. The rapid progress of nanotechnology allows the synthesis of many low-dimensional and nanoscale materials. This development provides with many new choices for field emission materials. For example, a new cathode for cold-field emission gun using a carbon-cone, supported by a carbon nanotube as the electron emitting tip has been developed  \cite{RefJ5,RefJ6,RefJ7,RefJ8}. The field emission process is commonly described by the well known Fowler-Nordheim  law  \cite{RefJ9}. The standard Fowler-Nordheim theory is based on quantum theory of tunneling \cite{RefJ10} and assumes that the emitter has planar geometry, while the most tips used in practice have curved emitting surfaces \cite{RefJ11} that is favorable to increase the acting field. The corresponding correction is provided by the factor, taking into account local field enhancement.

Another way to obtain effective electron emission is to employ the short laser pulses. For example, the ultrafast electron source most commonly used in different applications is based on electron emission induced by focusing an amplified femtosecond laser pulse on to a solid state surface \cite{RefJ12}. It should be noted that nonlinear photoemission (photoionization) of electrons from simple atomic systems \cite{RefJ13,RefJ14,RefJ15} has been observed and investigated in detail previously as fundamental physical phenomena \cite{RefJ16}.

Nonlinear ionization can be partially interpreted in the framework of  quasi-static model in which the bound electrons experience an effective potential formed by adding to the atomic potential the contribution due to the instantaneous laser electric field. The ionization rate may then be approximated by the averaged static-limit tunneling rate which can be easily calculated for hydrogenic type systems \cite{RefJ17}. Because of the exponential factor, tunneling occurs predominantly at the peaks of the electric field during the halfcycle when it lowers the potential barrier. As a result, the photoelectron wave packets are emitted in periodic bursts in time. In the energy domain, for sufficiently long pulses, this periodicity gives rise to the ATI spectrum.  ATI means that an atom may absorb more photons than are necessary for its ionization. This leads to photoelectron energies considerably higher than the photon energy. A more general quasi-static theory was developed to describe multiphoton ionization in the low-frequency limit  \cite{RefJ18,RefJ19,RefJ20}. The strong field approximation is made here, whereby it is assumed that an electron, after having being ionized, interacts only with the laser field and not with its parent core.

Despite achievements in the generation of high harmonics and attosecond pulses, atomic gases can not serve as a pulsed electron sources with high current densities. As such, it is preferable a process of non-linear electron emission from a solid surface under the laser pulse action.	Additional interest in the study of the interaction of laser light with the metal surface is caused by the development a number of laser applications related to monitoring, control and diagnostics of various nano-sized objects \cite{RefJ21,RefJ22,RefJ23}.

The study of the process of emission of electrons from a metal under the action of the laser pulse began almost immediately after the widespread use of lasers in experimental physics \cite{RefJ24}. In experiments with laser pulses sufficiently long duration electron emission was caused by heating the target, i.e., this process was limited to the thermal emission of electrons \cite{RefJ25,RefJ26}. With the advent of technology reducing the duration of the laser pulse, emission pattern has changed significantly. When the laser pulse duration, comparable to the electron-phonon relaxation, which, according to \cite{RefJ27,RefJ28} is  $\sim 1$ps, succeed avoid significant heating of the target even in the case of relatively high laser intensities. The emission of electrons from a metal under the influence of such pulse occurs mainly as a result of non-linear process without heating of the electron gas inside the target \cite{RefJ29,RefJ30}. The initial state of the whole ensemble of electrons does not have time to change significantly during the short laser pulse.

Ultrafast, laser-driven electron emission from metal nanostructures is of substantial current interest. The ultrashort coherent electron bunches produced are crucial to free electron lasers, laser acceleration of relativistic electrons, picosecond cathodoluminescence, and femtosecond electron diffraction. They would enable exciting technological developments such as four-dimensional time-resolved electron microscopy, spectroscopy, and holography. Intense multiphoton electron emission is observed from sharp metallic tips illuminated with even weak and short light pulses. Local field enhancement, evidenced by concurrent nonlinear light generation, confines the emission to the tip apex. The strong optical nonlinearity of the electron emission allows one to image the local optical field near a metallic nanostructure with a spatial resolution of a few tens of nanometers in a novel tip-enhanced electron emission microscope \cite{RefJ31}. It has been shown that a field emission tip electron source that is triggered with a femtosecond laser pulse can generate electron pulses even shorter than the laser pulse duration \cite{RefJ32}. It should be noted, that electron spectrum is not a constant depending only on the characteristics of the metal and light, but varies with the interaction duration in ultrashort time scales \cite{RefJ33}.

The photoelectric emission model for atoms in a strong monochromatic field \cite{RefJ18,RefJ19,RefJ20} was extended to the treatment of the field interacting with a metal \cite{RefJ34,RefJ35}. Such a model is based on the assumption that the photoemitted current density is a linear combination of many partial contributions, each characterized by its own order of nonlinearity \cite{RefJ36}.
According to the present time understanding, the nonlinear cold emission of electrons from the metal in a relatively weak low-frequency laser field usually is considered through two schemes. The first scheme is based on the time-dependent Schrodinger equation, the initial state of the electron in this case is described in the framework of the band model \cite{RefJ37}, and the final state is most often constructed as the Volkov function, including interaction with the electromagnetic field \cite{RefJ38}. Next, calculate the matrix elements corresponding to the transition amplitude, and on that basis electron energy spectrum is obtained. The second scheme is developed in the framework of the time-dependent density functional using the Kohn-Sham formalism \cite{RefJ39}. It is assumed that the electrons in the metal are independent, and they move in some effective potential produced by the crystal lattice ions in the presence of exchange interaction and electron correlation. The Kohn-Sham equations, describing the process of electron emission from a metal under the action of femtosecond laser pulse, can be solved only numerically, for example, using the Crank-Nicholson method. Emerging computational difficulties tightly associated with these schemes reduce their attractiveness. We construct a more simple model for nonlinear electron   cold emission from a metal under the femtosecond laser pulse action, that is devoid this shortcoming and gives quite convincing results when compared with experimental data.
\section{Theoretical model}
\label{sec:1}
The experimental data \cite{RefJ33} show a dependence on the laser polarization, indicating that the emission yield is about one order of magnitude larger for the $p$-polarized light electric field normal to the surface then for the $s$-polarization. This result could apparently suggest, according to the model proposed by Broudy \cite{RefJ40}, that the extraction mechanisms are directly related to the component of the electric field normal to the surface . Appropriate analytical one-dimensional model of electron emission from a metal surface, under the influence of both a dc electric field and laser field illumination with exact solution was derived to electron emission for arbitrary values of laser field, laser frequency, dc electric field, and metal work function \cite{RefJ41}. However, the considered electron emission from a metal-vacuum interface driven by a combination of a dc electric field and a laser field was limited to a single carrier frequency.

We  improve the model, describing the motion of electrons in a half-space under the influence of arbitrary pulsed laser field. The boundary conditions will ensure harmonization with the asymptotic behavior of the metal electronic states outside its surface. We write down the one-dimensional Schrodinger equation for an electron in the laser pulse field, described by vector potential $A(t)$ for the half-space outside the metal
\begin{equation} \label{eq1}	
i\frac{\partial \psi\left( x,t\right) }{\partial t}
=
\frac12 \left( \hat p + \frac1c A \left( t\right) \right) ^2
\psi\left( x,t\right).
\end{equation}
Here $\hat p$ is the momentum operator, $\psi\left( x,t\right)$ is a wave function.
We use the atomic system of units in which $m=\hbar=e=1$.

Consider laser field with a  strength $F\left( t\right)\ll1$, and assume that the external electromagnetic field weakly distorts an electron state  in the metal.
The initial states belonging to the solid target band structure are taken as exponentially decreasing into the vacuum. We note that in previous considerations the final states are Volkov states with momentum and energy in the vacuum half-space. The used final states were only defined in vacuum \cite{RefJ42}, and there was no extension of the approach to the states that are solutions of the one-dimension Schrodinger equation including the boundary conditions.

Under our assumptions, the wave function $\psi\left( x,t\right)$ satisfies the boundary conditions
\begin{eqnarray} \label{eq2}
\psi\left( 0,t\right)  =  B \exp \left( -i E_0 t\right), \nonumber \\
\psi_x^\prime\left( 0,t\right) = - \kappa B \exp \left( -i E_0 t\right).
\end{eqnarray}
Here $B$ is a normalization constant, $\kappa=\sqrt{|2 E_0|}$, $E_0$  is the energy of an electron inside the metal, prime in the Eq.~(\ref{eq2}) denotes the derivative with respect to the spatial variable.

We introduce the notation $v( t) =A\left( t\right) /c$ and write  Eq.~(\ref{eq1}) in the form
\begin{equation} \label{eq3}
i\frac{\partial \psi }{\partial t}
=
-\frac12 \frac{\partial ^2 \psi }{\partial x ^2}
-iv\left( t\right) \frac{\partial \psi }{\partial x }
+\frac{v^2\left( t\right)}{2} \psi.
\end{equation}
The solution of Eq.~(\ref{eq3}) with the boundary conditions Eq.~(\ref{eq2}) in the positive half-space $\left( x>0\right) $ can be found using the Laplace transform
$f\left( s,t\right) =\int\limits_{0}^{\infty}\exp\left( -sx\right)\psi\left( x,t\right) dx,$
where  $f\left( s,t\right)$  \cite{RefJ43} is transform  of function $\psi\left( x,t\right),$  $s$ is complex variable. In the transform space Eq.~(\ref{eq3}) takes the form
\begin{multline} \label{eq5}
i\frac{\partial f\left(s,t\right)  }{\partial t}
=
\frac12\left( -is+v \left( t\right) \right) ^2f\left(s,t\right)
\\ {} +
\left\lbrace   s \frac{\alpha\left( t\right)}{2} + \frac{\beta\left( t\right) }{2}  + i v \left( t\right)  \alpha\left( t\right)  \right\rbrace  .
\end{multline}
For more compact result representation we introduce notations:  $\alpha\left( t\right)=\psi\left( 0, t\right)$  and $\beta\left( t\right)=\psi_x^\prime\left( 0, t\right)$ . If you enter additional parameters for a classical electron displacement under the action of the laser field
\begin{equation} \label{eq6}
a( t,t_1) =\int\limits_{t_1}^{t} v(t_2) dt_2
\end{equation}
and the integral from the kinetic energy
\begin{equation} \label{eq7}
S( t,t_1) =\int\limits_{t_1}^{t} \frac{v^2(t_2)}{2} dt_2,
\end{equation}
solution of  Eq.~(\ref{eq5}) can be written as
\begin{multline} \label{eq8}
f(s,t)
=
-i\int\limits_{0}^{t}
\left\lbrace   s \frac{\alpha\left( t_1\right)}{2} + \frac{\beta\left( t_1\right) }{2}  + i v \left( t_1\right)  \alpha\left( t_1\right)  \right\rbrace
\\{}\times
\exp\left\lbrace i\left(  \frac{s^2\left( t-t_1\right)}{2} +i s a\left( t,t_1\right) -S\left( t,t_1\right)\right) \right\rbrace
dt_1.
\end{multline}
The replacing $s=ip$  in the Eq.~(\ref{eq8}) gives the solution of the original Eq.~(\ref{eq1}) in the momentum representation, and calculation the spectral distribution of the emitted electrons as a function of the pulse parameters is reduced to the integral
\begin{multline} \label{eq9}
f\left(p,T\right)
=
-iB\int\limits_{0}^{T}
\left\lbrace    \frac{ip}{2}-\frac{\kappa}{2}+iv(t_1) \right\rbrace \exp(-iE_0t_1)
\\{}\times
\exp\left\lbrace i\left(  -\frac{p^2\left( T-t_1\right)}{2} - p a\left( T,t_1\right) -S\left( T,t_1\right)\right) \right\rbrace
dt_1.
\end{multline}
Here  $T$  is the finish time a pulse action.
With the aid a known function  $f\left(p,T) \right) $, the electron energy spectrum can be written as
\begin{equation} \label{eq10}
\frac{dN}{dE}
=
\frac{\left| f\left(\sqrt{2 E},T\right) \right| ^2_{T=\infty}}{\sqrt{2 E}}.
\end{equation}
Eq. (9) describes a spectrum of emitted electrons for definite initial energy.  However, we need take into account that electrons in the metal are distributed over the initial energy.  According to the Sommerfeld theory number of electron states in unit volume of metal accounted per unite energy interval is determined as
\begin{equation} \label{eq11}
\frac{dn}{dE_0}
=
\frac{1}{\pi^2}\frac{ \sqrt{2 \left( E_0-U_0\right) } }
{1+\exp\left( \frac{E_0-E_f}{k \theta}\right) }.
\end{equation}
Here $\theta$  is metal temperature, $E_f$   is the Fermi energy, $U_0$ is energy corresponding to the bottom of the conduction band, $k$ is the Boltzmann constant.

After integration over initial electronic states energy and taking into account Eq.~(\ref{eq11}), we obtain the final equation that gives the energy spectrum of the emitted electrons
\begin{multline} \label{eq12}
\frac{dN_e}{dE}=
\frac{1}{\pi^2}\int\limits_{U_0}^{E_f}
\sqrt{\frac{E_0-U_0}{E}}
\left(1+ \exp\left( \frac{E_0-E_f}{k \theta}\right)\right) ^{-1}
\\{}\times
\left| f_{E_0}\left(\sqrt{2 E},T \right)\right|  ^2_{T=\infty} dE_0.
\end{multline}
The integration in Eq.~(\ref{eq12}) is carried out from the bottom of the metal conduction band $U_0$ to the energy $E_f$. At the room temperature, the population electronic states above the Fermi level can be neglected.

\section{Electron spectrum}
\label{sec:2}

The photoemission process depends on the intensity of the laser field, the carrier frequency, and the work function for a metal. The different features of  the photoemission can be a result of tunneling or multiphoton absorption. The switching from one to another type of photoemission is regulated by the Keldysh parameter  $\gamma=\sqrt{|E_f|/2U_p}$  \cite{RefJ18} that is the  combination of the metal work function $|E_f|$  and ponderomotive potential  $U_p$. If  $\gamma\gg 1,$ the tunneling time is large compared with the period of the field, and electron does not have enough time to leave the metal during the barrier penetration. This is the case of a relatively weak field, when multiphoton ionization dominates. In the opposite limit, when  $\gamma\ll 1,$ electron time passing under the barrier is small compared with the period of the field, and the efficient tunneling is realised. Our model is equally applicable to the description of both limit cases, and  the mode with $\gamma\sim 1$ too .

In the few-cycle regime, the electric field $F(t)$  should be written as $F(t)=f(t) \cos (\omega t+\varphi)$   where $f(t)$  denotes the pulse envelope. It is evident, the electric field as a detailed function of time depends on the absolute phase, although the envelope is the same for all pulses. It means, the electric field time variation in such a pulse depends on the phase of the carrier wave with respect to the envelope.

We specify the form of the electron energy spectrum produced in the process of nonlinear emission from the metal under the action of femtosecond laser pulses.
Let us assume the dependence of the laser field on time is expressed as the Gaussian envelope in the form
\begin{equation} \label{eq13}
F\left( t\right)
=
F_0 \exp \left( -\frac{t^2}{\tau ^2 }\right)
 \cos \left( \omega  t + \varphi \right),
\end{equation}
where $F_0$  is the amplitude of the laser field strength at the maximum,  $\omega$  is the  laser  carrier frequency, $\varphi$    is the carrier-envelope phase, $\tau$ is a parameter  of the pulse duration.

Increasing the field strength leads  to the greater yield of electrons according to rising photoemission probability and  stretching of the electron energy spectrum that means the more effective electron acceleration by the laser field. This is confirmed by experimental data as well as  numerical simulation results  for  a tungsten nano-nidle with a radius of curvature ~ 50 nm within our model, both presented in Fig. 1. Bottom-up solid curves corresponds the laser field strengths 6.9 V/nm and 8.7 V/nm. The analogous couple of the dashed curves reproduce  experimental data \cite{RefJ45} for the laser field strengths 4.32 V/nm and 4.98 V/nm.
The excitation of surface plasmon polaritons \cite{RefJ46} near apex needle  provides enhanced laser field due to superfocusing. The experimental values  of the laser field strengths used in the numerical simulations are taken into account with this amplification. It should be noted that the experimental conditions characterized by the presence of additional dc electric field with the strength 0.8 V/nm. This means, that, despite a certain difference between the experimental and theoretical parameters, the total effective field in both cases is comparable in magnitude. The results of numerical simulations, definitely show correspondence with the experimental dependencies.
\begin{figure}
	\begin{center}
		\includegraphics[width=8.5cm,keepaspectratio]{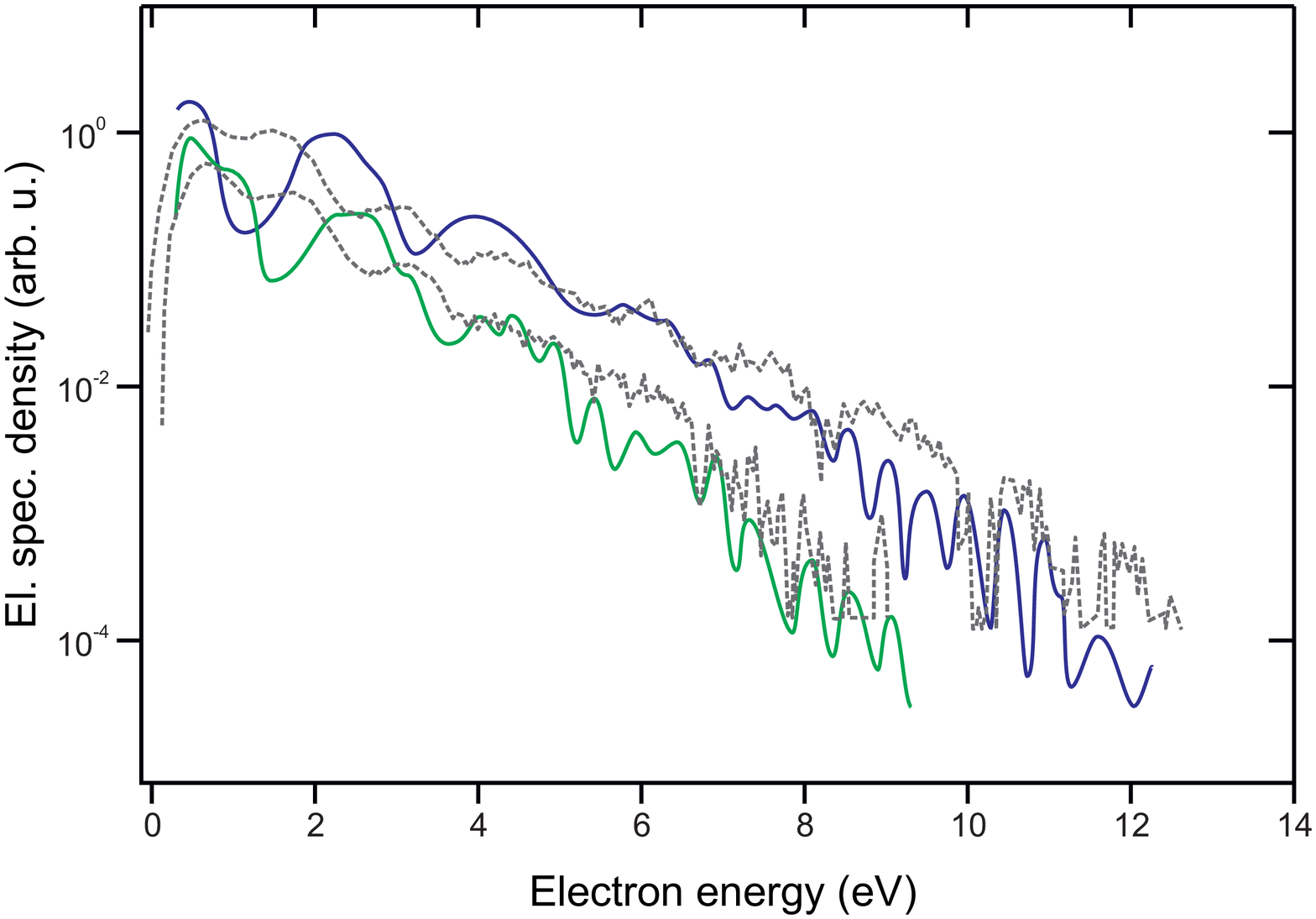}
		\caption{Theoretical and experimental photoelectron spectrums with averaging over  carrier-envelope phase for tungsten target: $\lambda = 800$~nm  and $\tau = 6.5$~fs.}
		\label{fig1}
	\end{center}
\end{figure}

For the results of numerical simulations, shown in Fig. ~\ref{fig1}, the Keldysh parameter ranges from 1.8 to 2.3, i.e., the non-linear emission is intermediate between the multiphoton emission and the tunneling, closer to the second mode. This is confirmed by existence of  characteristic peaks in the electron spectrum. The minimum number of photons needed for above-threshold electron emission is $K=\langle|E_f|/\omega\rangle+1$ , where  $\langle\ldots\rangle$ is the integer part of a number.  Since the electron work function for tungsten nanotip  is near $ 4$ eV, in this experiment $K=3$. The position of the first peak in Fig. 1 corresponds to this number. Subsequent peaks with a much lower yield of electrons are separated by almost equidistant gaps $\sim\omega .$
The spectrum of applied laser light is broad enough when we have a very short pulse, and some contribution to the photoemission  is produced by electrons having an initial energy below the Fermi level. As a result, the energy spectrum of the emitted electrons is broadened, and the gap between peaks in the energy spectrum is not strictly constant.
All the dependencies have a maximum near three photon ionization threshold. Characteristic feature is the presence of a plateau, following the peak.

Then, at a certain electron energy value $E_{cut}$  (cutoff energy) sharp decline in spectrum dependence takes place.
With the growth of the amplitude of the laser field, when all other parameters are fixed, the value of the cutoff energy, the part of electrons with high energy and the total number of emitted electrons are increased.  The cutoff energy increases proportionally  ponderomotive potential, i.e., the average electron vibration energy.

\begin{figure}
	\begin{center}
		\includegraphics[width=8.5cm,keepaspectratio]{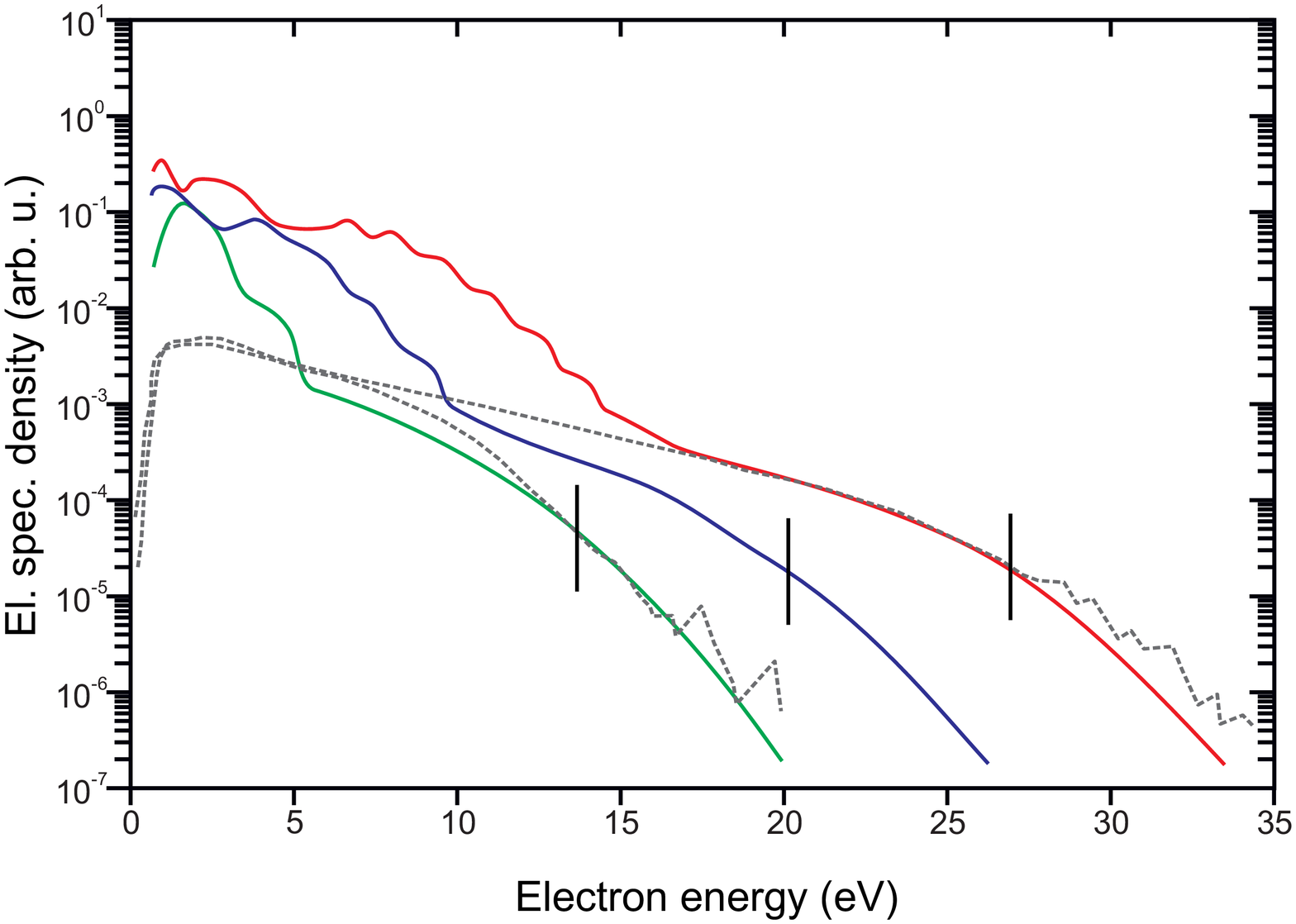}
		\caption{Theoretical and experimental photoelectron spectrums with averaging over  carrier-envelope phase for gold target: $F_0 =14$ V/nm, $\tau  = 150$ fs. Bottom-up solid curves are plotted for laser wavelengths 960 nm, 1030 nm, and 1100 nm. Dashed curves shows the experimental data \cite{RefJ47} for laser wavelengths 960 nm and 1100 nm.}
		\label{fig2}
	\end{center}
\end{figure}

A comprehensive study of photoemission strong-field was carried \cite{RefJ47} using gold nanotip, the tip  curvature radius of which was ~ 10 nm. The experimental setup allowed to obtain laser pulses of $\sim 150$ fs with different intensities and wavelengths of carrier. In the control process of the total photoelectron yield,  can be maintain constant value the electric  field amplitude near nanotip surface. Due to this, it is possible to obtain experimentally photoelectron spectrum at various wavelengths of the laser radiation, fixing the other parameters. In Fig. ~\ref{fig2} dashed lines represent some experimental data, corresponding to different values of wavelengths, as well as solid lines show the result of their theoretical recovery in our model with averaging over the phase $\varphi.$ Vertical bars indicate cutoff energies.
The variation in shape the curves shown in Fig. 1,2 reflects the change in photoemission  mode from the intermediate between the multiphoton  and the tunneling emission to  the more pure tunnel regime. Note that direct interpretation of the experimental results is seriously complicated by the uncertainty of the  experimental value of the dc electric field \ cite \cite{RefJ47}.

The carrier phase control with respect to the envelope demonstrates the dependence on phase the spectrum of electron emission from nanoscale needle under the action of laser pulse \cite{RefJ48}. This dependence in the experiments with nanotips is even more pronounced than in similar experiments with the atomic gases \cite{RefJ49}. The gas gives the blur effect associated with the  large size of the emission space and the resulting spatial inhomogeneity of the laser radiation. In contrast, electrons are emitted from nanotip in a small region with a characteristic size of not more than 10 nm, where the  laser field  strength can be considered as constant.

\begin{figure}
	\begin{center}
		\includegraphics[width=8.5cm,keepaspectratio]{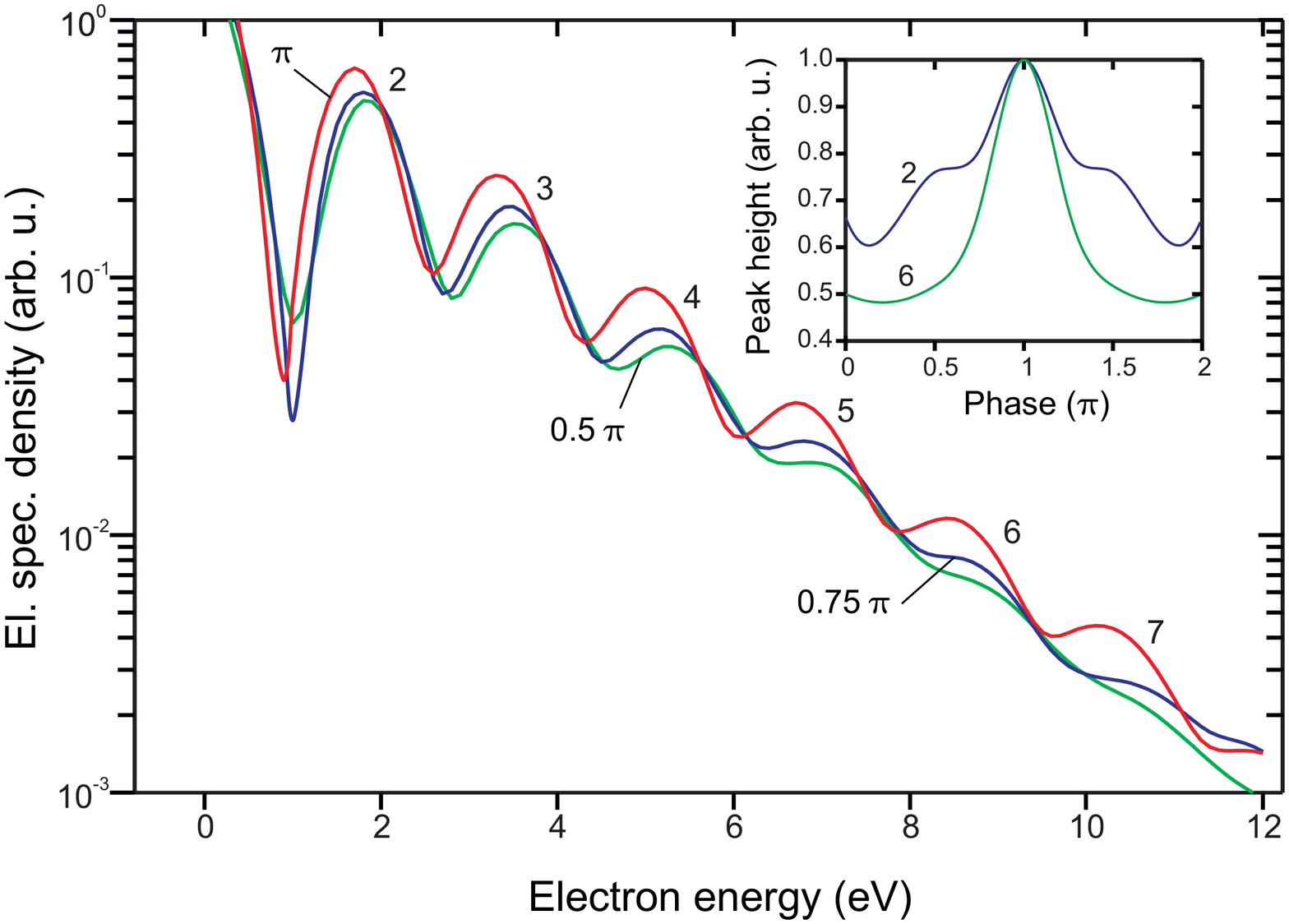}
		\caption{ Photoelectron spectrums for different carrier-envelope phases:  $F_0=8.7$~nm, $\lambda=800$~nm, $\tau=6.5$~fs.
The phases for different curves are indicated in the figure.}
		\label{fig3}
	\end{center}
\end{figure}

Fig. ~\ref{fig3} shows the results of calculation of the spectrum of emitted electrons at different values of the phase. The clearly-visible changes in the peak heights are presented on all three curves. These dependences maith be due to the variation in the degree of quantum interference between the electron wave packets resulting from rescattering on the metal surface during different optical cycles. For example, when $\varphi=\pi$ a large peak height indicates that in at least two wave packets give a contribution to spectrum. At increasing the electron energy the phase effects are manifested to a greater extent. To demonstrate this, the inset in Fig. 3 shows the normalized to unity at a maximum  values altitudes of  the second and sixth peaks  as a function of phase.These dependences qualitatively consistent with experimental data \cite{RefJ48}.
For long multicycle pulses the role of phase in the electron spectrum formation is not essential. Thus, we can conclude that the phase effects are essential only for laser pulses with a few femtoseconds duration.

Earlier, for atoms in gas target using laser pulses having duration of approximately 6 fs \cite{RefJ46a}, the similar effect was demonstrated for ATI photoelectrons. It has been reported about sub-femtosecond control of the electron emission in ATI of the noble gases Ar, Kr and Xe in intense, few-cycle laser fields  \cite{RefJ48}.

\section{Conclusions}
\label{sec:3}

We formulated the one-dimensional model for the nonlinear field electron emission from a metal surface under ultrashort laser pulse action. In the framework of this approach we investigated the influence the laser pulse parameters on the energy spectrum of the emitted electrons. Comparison the results of the numerical simulations with the experimental data show that the model reproduces the general  spectrum form. Selecting target as nanotip, that focuses the laser pulse, provides a significant increase the laser field near surface of the target  due to surface plasmon polariton  superfocusing \cite{RefJ50,RefJ51}. The magnitude of the field strength is given taking into account the effect of its gain.  Arising surface plasmon resonance is a collective oscillation of conduction band electrons that typically occurs at optical frequencies in noble metals \cite{RefJ53}. For a short amount of time, these electrons oscillate in phase with each other and create a strongly enhanced electric field at  the metal/vacuum interface \cite{RefJ54}.

We obtained rather good agreement of numerical simulation results with some features of the experimental electron spectrum. The cutoff energy value, the general form of electron energy distribution, and the curve slope in the central part of the spectrum are reproduced. However, the developed theoretical model takes into account not all of the factors that influence the experimental photoemission process. Because of this, there is a certain discrepancy between the results obtained on its basis, with the experimental data. Experimental measurements were carried out in the presence of a significant external dc electric field. Its presence increases the output of electrons, and should change their energy spectrum, which is not taken into account in our calculations. It should also be noted the possible role of the Stark effect, which increases the electron work function. For a high-intensity laser field, this can lead to some shift of the whole spectrum is not taken into account in our model.
Our theoretical calculations reproduce well enough the high-energy part of the spectrum, however, for low electron energies the distinction with experimental data is observed. Most notably, it is shown in comparison with experiment with golden nanotip in which  dc electric field, near the tip, and its role in nonlinear electron emissions need to be studied in more details.

Changing the wavelength of the laser carrier can vary significantly the very character of interaction with the field emitted electrons. This is due to the fact that the gain region of the laser field near nanotip has a finite size, its order of magnitude comparable to the radius of curvature of a tip. In increasing wavelength, electron oscillation amplitude can exceed the size of this region (sub-cycle regime), an electron can leave it, and further movement will occur in the weak field. For  theoretical description of this case is suitable semiclassical two-step model  [47]. In the framework of the model in the first step in a quasi-static approximation   the amount of emitted electrons is determined on the basis of the Fowler-Nordheim  theory. In the second step,  electron dynamics is calculated using classical laws of motion. This model is applicable when the sub-cycle  mode and practically useless for multi-cycle regime, i.e., when the amplitude of the electron oscillations is much smaller then the spatial size of local field enhancement near the nanotip. In contrast to the two stage model, our model is suitable for the  multi-cycle regime and thus complements semiclassical description photoemission process.
To eliminate the remaining discrepancy between theory and experiment additional researchers are  highly desirable.

This work was supported by RFBR (grant No 16-32-00255) and Ministry of
Science and Education RF (Project No 2014/19-2881).

\end{document}